\documentclass[preprint,prb,superscriptaddress]{revtex4}
\usepackage{amsfonts}

\input{tcilatex}

\begin{document}

\preprint{Cond-Mat}
\title[Short title for running header]{$^{7}$Li NMR, magnetic
susceptibility, and heat capacity \ studies on the triangular lattice system
LiCrO$_{2}$.}
\author{L. K. Alexander}
\affiliation{Department of Physics, Indian Institute of Technology, Mumbai 400076, India}
\author{N. B\"{u}ttgen}
\affiliation{Experimentalphysik V, Elektronische Korrelationen und Magnetismus, Institut f%
\"{u}r Physik, Universit\"{a}t Augsburg, D-86135 Augsburg, Germany.}
\author{R. Nath}
\affiliation{Experimentalphysik V, Elektronische Korrelationen und Magnetismus, Institut f%
\"{u}r Physik, Universit\"{a}t Augsburg, D-86135 Augsburg, Germany.}
\author{A. V. Mahajan}
\affiliation{Department of Physics, Indian Institute of Technology, Mumbai 400076, India}
\author{A. Loidl}
\affiliation{Experimentalphysik V, Elektronische Korrelationen und Magnetismus, Institut f%
\"{u}r Physik, Universit\"{a}t Augsburg, D-86135 Augsburg, Germany.}

\begin{abstract}
We report $^{7}$Li NMR, magnetic susceptibility, and heat capacity
measurements on the triangular lattice Heisenberg antiferromagnet compound
LiCrO$_{2}$. We find that in contrast to NaCrO$_{2}$, magnetic properties of
LiCrO$_{2}$ have a more pronounced three dimensional character with sharp
anomalies in the temperature variation of the $^{7}$Li NMR intensity and the
NMR spin-lattice relaxation rate 1/T$_{1}$. \ From heat capacity
measurements we find that the total entropy related to the magnetic
transition is in agreement with expectations. \ However, we find a
significant contribution to the magnetic entropy in the range from the
ordering temperature $T_{\text{N}}$ to nearly $4T_{\text{N}}$. \ This
suggests the existence of magnetic correlations at temperatures well above $%
T_{\text{N}}$ which might be due to the frustrated nature of the system.
Based on the temperature dependence of 1/T$_{1}$, we discuss the possible
occurrence of a Kosterlitz-Thouless-Berezinskii transition taking place at $%
T_{\text{KTB}}$ = $55$ K in LiCrO$_{2}$. Lithium depletion has no
significant effect on the magnetic properties and the behaviour of Li$_{0.5}$%
CrO$_{2}$ is nearly unchanged from that of LiCrO$_{2}$.
\end{abstract}

\volumeyear{year}
\volumenumber{number}
\issuenumber{number}
\eid{identifier}
\date[Date text]{date}
\received[Received text]{date}
\revised[Revised text]{date}
\accepted[Accepted text]{date}
\published[Published text]{date}
\startpage{1}
\endpage{2}
\maketitle

\section{\textbf{INTRODUCTION}}

Geometrically frustrated systems are being widely studied due to the
possibility of unconventional ground states and their susceptibility to weak
perturbations. The two-dimensional triangular lattice Heisenberg
antiferromagnet (TLHAF) is a prominent example where exotic magnetic
phenomena have been observed. Specifically, while Na$_{x}$CoO$_{2}$.$1.3$H$%
_{2}$O is superconducting $(0.25<x<0.33)$, the unhydrated Na$_{x}$CoO$_{2}$
also shows a gamut of magnetic transitions\cite{naxcoo2 supercond,naxcoo2
sugiyama,naxcoo2 mlfoo}. Presumably with the above background, Olariu 
\textit{et al}.\cite{nacro2 mendels}\textit{\ }investigated NaCrO$_{2}$ by
nuclear magnetic resonance (NMR) and muon spin resonance ($\mu $SR)
experiments (in addition to bulk probes) and thereby concluded to have
observed for the first time an "extended fluctuating regime" in TLHAF.
Recent reports on NiGa$_{2}$S$_{4}$ (ref. 5) suggest an unconventional
magnetic ground state originating from magnetic correlations which extend
beyond nearest neighbours.

In this paper we present a detailed study of the magnetism and heat capacity
of LiCrO$_{2}$ using both macroscopic and local probes. Further, in view of
the important role played by sodium ions in Na$_{x}$CoO$_{2}$, measurements
on the delithiated compound Li$_{0.5}$CrO$_{2}$ are also presented. Previous
work on LiCrO$_{2}$ has established that it possess $\alpha -$NaFeO$_{2}$
type structure comprising two-dimensional triangular chromium layers \cite%
{licro2 tauber}. Magnetic susceptibility data \textit{\ }\cite{licro2
tauber,licro2 delmas}\textit{\ }indicate a transition at around $62$ K.
Published susceptibility data could be fitted with a Curie-Weiss law only
above $450$ K and the paramagnetic Curie temperature is reported in the
range $570$ K to $700$ K\cite{licro2 tauber,licro2 delmas,licro2 moreno}.
Neutron scattering and Raman scattering studies\cite{licro2 sobeyroux,licro2
kadowaki,licro2 suzuki} suggest a $3D$ character to the magnetic
correlations below the transition temperature ($T_{\text{N}}$). Above $T_{%
\text{N}}$, measurements indicate a wide region showing short-range magnetic
correlations\cite{licro2 moreno}. There are no NMR\ studies or specific heat
measurements reported on the compound till date.

Herein, we report bulk susceptibility, $^{7}$Li NMR\ studies, and specific
heat measurements carried out on Li$_{x}$CrO$_{2}$. $(x=1,0.5).$We found
that the $^{7}$Li NMR spectral intensity of the title compounds drops
precipitously at $T_{\text{N}}$ unlike in NaCrO$_{2}$ where the intensity
drops progressively in a 10 K range below $T_{\text{N}}$.\ The "extended
fluctuating regime" observed in NaCrO$_{2}$ does not appear pronounced in
LiCrO$_{2}$. Our heat capacity measurements exhibit a peak at the magnetic
ordering temperature. Both NMR and specific heat studies suggest that
magnetic correlations develop well above $T_{\text{N}}$. The rest of the
paper is organised as follows. Section II contains the experimental details,
wherein synthesis and structural characterisation of Li$_{x}$CrO$_{2}$,
experimental set-up and parameters for various other measurements such as
bulk magnetic susceptibility, specific heat, and NMR are included. In
section III, we present the results of our measurements which is followed by
a comprehensive discussion in section IV.

\section{\textbf{EXPERIMENTAL DETAILS}}

Polycrystalline LiCrO$_{2}$ was synthesised by solid-state reaction of a
stoichiometric mixture of Li$_{2}$CO$_{3}$ (Aldrich, 99.99 \%) and Cr$_{2}$O$%
_{3}$ (Aldrich, 98.0\%). The mixture was fired at 800$%
%TCIMACRO{\U{b0}}%
%BeginExpansion
{{}^\circ}%
%EndExpansion
$C for 48 hours with one intermediate grinding. X-ray diffraction experiment
was carried out using PANalytical X'pert PRO powder diffractometer with
X'celerator detector. The diffractometer uses a Cu-K$_{\alpha }$ target ($%
\lambda _{av}=1.54182$ \AA ). Our results (see Fig. 1(a)) indicate single
phase formation and the peaks can be indexed using the trigonal space group, 
$\mathit{R}\overline{3}m.$Using a least-squares method, the lattice
parameters were found to be $a=2.8949(12)$ \AA $,$ $b=2.8949(12)$ \AA , and $%
c=14.3886(98)$ \AA . As shown in a sketch of the LiCrO$_{2}$ crystal
structure (Fig. 2), Cr layers have a triangular configuration. In the
structure, oxygen ions form ABCABC type of stacking layers. Also, the edge
shared Cr octahedral layers are separated by Li ions. The crystal structure
makes a strong case for $2D$ triangular lattice Heisenberg antiferromagnets.
\ \ \ \ \ \ \ \ \ \ \ \ \ \ \ \ \ \ \ \ \ \ \ \ \ \ \ \ \ \ \ \ \ \ \ \ \ \
\ \ \ \ \ \ \ \ \ \ \ \ \ \ \ \ \ \ \ \ \ \ \ \ \ \ \ \ \ \ \ \ \ \ \ \ \ \
\ \ \ \ \ \ \ \ \ \ \ \ \ \ \ \ \ \ \ \ \ \ \ \ \ \ \ \ \ \ \ \ \ \ \ \ \ \
\ \ \ \ \ \ \ \ \ \ \ \ \ \ \ \ \ \ \ \ \ \ \ \ \ \ \ \ \ \ \ \ \ \ \ \ \ \
\ \ \ \ \ \ \ \ \ \ \ \ \ \ \ \ \ \ \ \ \ \ \ \ \ \ \ \ \ \ \ \ \ \ \ \ \ \
\ \ \ \ \ \ \ \ \ \ \ \ \ \ \ \ \ \ \ \ \ \ \ \ \ \ \ \ \ \ \ \ \ \ \ \ \ \
\ \ \ \ \ \ \ \ \ \ \ \ \ \ \ \ \ \ \ \ \ \ \ \ \ \ \ \ \ \ \ \ \ \ \ \ \ \
\ \ \ \ \ \ \ \ \ \ \ \ \ \ \ \ \ \ \ \ \ \ \ \ \ \ \ \ \ \ \ \ \ \ \ \ \ \
\ \ \ \ \ \ \ \ \ \ \ \ \ \ \ \ \ \ \ \ \ \ \ \ \ \ \ \ \ \ \ \ \ \ \ \ \ \
\ \ \ \ \ \ \ \ \ \ \ \ \ \ \ \ \ \ \ \ \ \ \ \ \ \ \ \ \ \ \ \ \ \ \ \ \ \
\ \ \ \ \ \ \ \ \ \ \ \ \ \ \ \ \ \ \ \ \ \ \ \ \ \ \ \ \ \ \ \ \ \ \ \ \ \
\ \ \ \ \ \ \ \ \ \ \ \ \ \ \ \ \ \ \ \ \ \ \ \ \ \ \ \ \ \ \ \ \ \ \ \ \ \
\ \ \ \ \ \ \ \ \ \ \ \ \ \ \ \ \ \ \ \ \ \ \ \ \ \ \ \ \ \ \ \ \ \ \ \ \ \
\ \ \ \ \ \ \ \ \ \ \ \ \ \ \ \ \ \ \ \ \ \ \ \ \ \ \ \ \ \ \ \ \ \ \ \ \ \
\ \ \ \ \ \ \ \ \ \ \ \ \ \ \ \ \ \ \ \ \ \ \ \ \ \ \ \ \ \ \ \ \ \ \ \ \ \
\ \ \ \ \ \ \ \ \ \ \ \ \ \ \ \ \ \ \ \ \ \ \ \ \ \ \ \ \ \ \ \ \ \ \ \ \ \
\ \ \ \ \ \ \ \ \ \ \ \ \ \ \ \ \ \ \ \ \ \ \ \ \ \ \ \ \ \ \ \ \ \ \ \ \ \
\ \ \ \ \ \ \ \ \ \ \ \ \ \ \ \ \ \ \ \ \ \ \ \ \ \ \ \ \ \ \ \ \ \ \ \ \ \
\ \ \ \ \ \ \ \ \ \ \ \ \ \ \ \ \ \ \ \ \ \ \ \ \ \ \ \ \ \ \ \ \ \ \ \ \ \
\ \ \ \ \ \ \ \ \ \ \ \ \ \ \ \ \ \ \ \ \ \ \ \ \ \ \ \ \ \ \ \ \ \ \ \ \ \
\ \ \ \ \ \ \ \ \ \ \ \ \ \ \ \ \ \ \ \ \ \ \ \ \ \ \ \ \ \ \ \ \ \ \ \ \ \
\ \ \ \ \ \ \ \ \ \ \ \ \ \ \ \ \ \ \ \ \ \ \ \ \ \ \ \ \ \ \ \ \ \ \ \ \ \
\ \ \ \ \ \ \ \ \ \ \ \ \ \ \ \ \ \ \ \ \ \ \ \ \ \ \ \ \ \ \ \ \ \ \ \ \ \
\ \ \ \ \ \ \ \ \ \ \ \ \ \ \ \ \ \ \ \ \ \ \ \ \ \ \ \ \ \ \ \ \ \ \ \ \ \
\ \ \ \ \ \ \ \ \ \ \ \ \ \ \ \ \ \ \ \ \ \ \ \ \ \ \ \ \ \ \ \ \ \ \ \ \ \
\ \ \ \ \ \ \ \ \ \ \ \ \ \ \ \ \ \ \ \ \ \ \ \ \ \ \ \ \ \ \ \ \ \ \ \ \ \
\ \ \ \ \ \ \ \ \ \ \ \ \ \ \ \ \ \ \ \ \ \ \ \ \ \ \ \ \ \ \ \ \ \ \ \ \ \
\ \ \ \ \ \ \ \ \ \ \ \ \ \ \ \ \ \ \ \ \ \ \ \ \ \ \ \ \ \ \ \ \ \ \ \ \ \
\ \ \ \ \ \ \ \ \ \ \ \ \ \ \ \ \ \ \ \ \ \ \ \ \ \ \ \ \ \ \ \ \ \ \ \ \ \
\ \ \ \ \ \ \ \ \ \ \ \ \ \ \ \ \ \ \ \ \ \ \ \ \ \ \ \ \ \ \ \ \ \ \ \ \ \
\ \ \ \ \ \ \ \ \ \ \ \ \ \ \ \ \ \ \ \ \ \ \ \ \ \ \ \ \ \ \ \ \ \ \ \ \ \
\ \ \ \ \ \ \ \ \ \ \ \ \ \ \ \ \ \ \ \ \ \ \ \ \ \ \ \ \ \ \ \ \ \ \ \ \ \
\ \ \ \ \ \ \ \ \ \ \ \ \ \ \ \ \ \ \ \ \ \ \ \ \ \ \ \ \ \ \ \ \ \ \ \ \ \
\ \ \ \ \ \ \ \ \ \ \ \ \ \ \ \ \ \ \ \ \ \ \ \ \ \ \ \ \ \ \ \ \ \ \ \ \ \
\ \ \ \ \ \ \ \ \ \ \ \ \ \ \ \ \ \ \ \ \ \ \ \ \ \ \ \ \ \ \ \ \ \ \ \ \ \
\ \ \ \ \ \ \ \ \ \ \ \ \ \ \ \ \ \ \ \ \ \ \ \ \ \ \ \ \ \ \ \ \ \ \ \ \ \
\ \ \ \ \ \ \ \ \ \ \ \ \ \ \ \ \ \ \ \ \ \ \ \ \ \ \ \ \ \ \ \ \ \ \ \ \ \
\ \ \ \ \ \ \ \ \ \ \ \ \ \ \ \ \ \ \ \ \ \ \ \ \ \ \ \ \ \ \ \ \ \ \ \ \ \
\ \ \ \ \ \ \ \ \ \ \ \ \ \ \ \ \ \ \ \ \ \ \ \ \ \ \ \ \ \ \ \ \ \ \ \ \ \
\ \ \ \ \ \ \ \ \ \ \ \ \ \ \ \ \ \ \ \ \ \ \ \ \ \ \ \ \ \ \ \ \ \ \ \ \ \
\ \ \ \ \ \ \ \ \ \ \ \ \ \ \ \ \ \ \ \ \ \ \ \ \ \ \ \ \ \ \ \ \ \ \ \ \ \
\ \ \ \ \ \ \ \ \ \ \ \ \ \ \ \ \ \ \ \ \ \ \ \ \ \ \ \ \ \ \ \ \ \ \ \ \ \
\ \ \ \ \ \ \ \ \ \ \ \ \ \ \ \ \ \ \ \ \ \ \ \ \ \ \ \ \ \ \ \ \ \ \ \ \ \
\ \ \ \ \ \ \ \ \ \ \ \ \ \ \ \ \ \ \ \ \ \ \ \ \ \ \ \ \ \ \ \ \ \ \ \ \ \
\ \ \ \ \ \ \ \ \ \ \ \ \ \ \ \ \ \ \ \ \ \ \ \ \ \ \ \ \ \ \ \ \ \ \ \ \ \
\ \ \ \ \ \ \ \ \ \ \ \ \ \ \ \ \ \ \ \ \ \ \ \ \ \ \ \ \ \ \ \ \ \ \ \ \ \
\ \ \ \ \ \ \ \ \ \ \ \ \ \ \ \ \ \ \ \ \ \ \ \ \ \ \ \ \ \ \ \ \ \ \ \ \ \
\ \ \ \ \ \ \ \ \ \ \ \ \ \ \ \ \ \ \ \ \ \ \ \ \ \ \ \ \ \ \ \ \ \ \ \ \ \
\ \ \ \ \ \ \ \ \ \ \ \ \ \ \ \ \ \ \ \ \ \ \ 

Li$_{0.5}$CrO$_{2}$ was prepared by extraction of lithium from LiCrO$_{2}$
using chemical methods\cite{licoo2 manthiram}. The extraction was carried
out by stirring LiCrO$_{2}$ powders in an aqueous solution of an oxidising
agent, Na$_{2}$S$_{2}$O$_{8}$ (Thomas Baker, 99.0 \%) for two days with help
of a magnetic stirrer. The amount of Na$_{2}$S$_{2}$O$_{8}$ required for the
delithiation reaction was decided based on our experience with LiCoO$_{2}$ 
\cite{lixcoo2 paper}. The delithiation reaction for synthesis of Li$_{0.5}$%
CrO$_{2}$ can be summed up as:

$2$LiCrO$_{2}$ + $0.5$ Na$_{2}$S$_{2}$O$_{8}$ $\longrightarrow $ $2$ Li$%
_{0.5}$CrO$_{2}$ + $0.5$ Na$_{2}$SO$_{4}$ + $0.5$ Li$_{2}$SO$_{4}$

The product was filtered, washed repeatedly - first with water and finally
with acetone, and air-dried. In order to estimate the level of delithiation
in LiCrO$_{2}$, $^{7}$Li NMR\ spectral intensities were compared. Here, the
NMR spectra were measured in the un-delithiated and the delithiated samples
under identical conditions and normalised by the respective sample masses.
Based on this, Li$_{0.5}$CrO$_{2}$ should be written as Li$_{(0.51\pm 0.02)}$%
CrO$_{2}$. XRD pattern of Li$_{0.5}$CrO$_{2}$ was found to match exactly
with that of LiCrO$_{2}$(Fig.1(b)). \ Lattice parameters of Li$_{0.5}$CrO$%
_{2}$ do not vary significantly from those of LiCrO$_{2}$ quoted in the
above paragraph.

Macroscopic magnetic properties were measured using a superconducting
quantum interference device (SQUID) magnetometer (MPMS5, Quantum Design) and
a vibrating sample magnetometer (VSM) from Quantum Design. Magnetization ($M$%
) was measured as a function of temperature ($T$) in an applied field ($H$)
of $1$ T, in a temperature range $2$ K-$300$ K. The $M$-$H$ isotherms were
measured at $280$ K, $200$ K, $150$ K, and $50$ K. None of these $M$-$H$
plots (not shown) exhibited any signs of hysteresis. Ferromagnetic
impurities estimated using the plots were found to be negligible. Zero-field
cooled (ZFC) and field cooled (FC) magnetization measurements were carried
out on both the samples. In this experiment, sample was first cooled in zero
applied field to $2$ K and then the measurements were taken from $2$ K to $%
300$ K in a field of $50$ G (ZFC data). Whereas for the FC measurements,
sample was cooled in an applied field of $50$ G. Then in the same field,
data were collected in the temperature $2$ K to $300$ K.

Solid-state NMR\ is an efficient local probe of magnetic properties. This
technique allows one to study static magnetic correlations and low-energy
spin excitations. Our NMR experiments were carried out using a home-built
NMR\ spectrometer. \ We used an Oxford Instruments (sweepable)
superconducting magnet and a variable temperature insert. In the
experiments, $^{7}$Li nuclei (nuclear spin $I$\textit{\ }=$3/2$\ and
gyromagnetic ratio $\gamma $/2$\pi $ = $16.546$ MHz/Tesla ) were probed
using pulsed NMR technique. Measurements were carried out in the temperature
range, $2$ K - $300$ K. Spectra were obtained employing Fourier transform
(FFT) technique at a fixed field of $1.089$ T. At low temperatures (below $10
$ K), since the spectra were too broad for an accurate FFT experiment, field
sweep measurement at a frequency of $18$ MHz were carried out to supplement
the spectral data. In order to check the field dependence for the NMR\
characteristics of the compound, measurement were also done at a higher
applied field; by sweeping the field at $71$ MHz (applied field $H\approx 4.3
$ T). It also allows to investigate the field dependence of the extended
fluctuation regime, if any. An aqueous solution of LiCl was used as the
diamagnetic reference. NMR\ spin echoes were obtained using $\pi /2-\tau
-\pi $ pulse sequences. NMR experiments at two different frequencies were
performed in order to check if there is any field dependence for the NMR
characteristics of the compounds. At $18$ MHz, $\pi /2$ pulse widths were
typically in the range $1-8$ $\mu s$ for LiCrO$_{2}$ and Li$_{0.5}$CrO$_{2}$%
. At $71$ MHz, typical $\pi /2$ pulse widths were $4$ $\mu s$. For
spin-lattice relaxation measurements, inversion recovery method was
employed. The spin-spin relaxation rate (1/T$_{2}$) was determined by
measuring the decay of echo intensity as a function of the separation
between the $\pi /2$ and $\pi $ pulse.

Specific heat measurements were carried out on LiCrO$_{2}$ in the
temperature range $3$ K to $300$ K using a Quantum Design PPMS system. LiCoO$%
_{2}$ is a non- magnetic equivalent\cite{lixcoo2 paper} of LiCrO$_{2}$
having the same crystal structure. Therefore, in order to extract the
magnetic part of specific heat of LiCrO$_{2}$, isostructural LiCoO$_{2}$ was
measured. For this purpose, LiCoO$_{2}$ was prepared as reported elsewhere%
\cite{lixcoo2 paper}.

\section{\textbf{RESULTS}}

\subsection{\textbf{Bulk magnetic susceptibility}}

The temperature dependence of bulk magnetic susceptibility $\chi (T)$($=M/H$%
) for both LiCrO$_{2}$ and Li$_{0.5}$CrO$_{2}$ are shown in figure. 3. In
both the samples, one can see that $\chi $ increases with decreasing
temperature till a broad maximum is reached at about $65$ K and below this
temperature $\chi $ is nearly constant. In line with the reported results%
\cite{licro2 tauber,licro2 delmas,licro2 moreno}, we found that the
susceptibility data of LiCrO$_{2}$ do not obey the Curie-Weiss law below $%
300 $ K. A negligible low-temperature Curie tail implies the presence of
very low amounts of paramagnetic impurities and the high quality of our
samples. The $65$ K anomaly in LiCrO$_{2}$ has been associated with \
antiferromagnetic (AF) order\cite{licro2 sobeyroux,licro2 kadowaki}. A
nearly constant (but somewhat higher) value of susceptibility for Li$_{0.5}$%
CrO$_{2}$ compared to LiCrO$_{2}$ through the entire temperature region
measured and a slightly enhanced low-temperature Curie-tail appears to
indicate that delithiation has generated small amounts of paramagnetic
impurities which remained even after repeated washing. Figure 4 shows ZFC
and FC magnetisation data for LiCrO$_{2}$. There, one observes a splitting
of the ZFC and FC curves at around $240$ K. This history dependent
magnetization for LiCrO$_{2}$ has not been reported till now. In the case of
Li$_{0.5}$CrO$_{2}$, as well, we observed roughly the same features as that
of LiCrO$_{2}$. We checked for the possible origin of this ZFC-FC splitting
from any unnoticed impurities (in spite of the negligible Curie tail, and
the single phase x-ray diffraction spectrum), especially oxides of chromium.
CrO$_{2}$ has a ferromagnetic transition temperature of $400$ K. Cr$_{2}$O$%
_{3}$ and Cr$_{2}$O$_{5}$ have antiferromagnetic transitions at $307$ K and $%
125$ K, respectively. This practically rules out impurities as a cause for
the $240$ K anomaly. An intrinsic reason for the history dependent
magnetisation observed in Li$_{x}$CrO$_{2}$ is then indicated.

\subsection{\textbf{NMR measurements}}

\subsubsection{ \textbf{Spectra}}

As seen in Fig. 2, the Li nuclei can have a hyperfine coupling (presumably
via the oxygen ions) with the magnetic Cr$^{3+}$ ions which are present in
triangular planes above and below the Li planes. The spin susceptibility of
the Cr$^{3+}$ ions $\chi _{spin}(T)$ gives rise to a shift of the $^{7}$Li
resonance $K(T)$ following Eq. (1):

\begin{equation}
N_{A}\mu _{B}K(T)=A_{hf}\chi _{spin}(T)
\end{equation}%
where $N_{A}$ is the Avogadro number, $\mu _{B}$ is the Bohr magneton, and $%
A_{hf}$ is the hyperfine coupling constant. In both LiCrO$_{2}$ and Li$_{0.5}
$CrO$_{2}$ at both the frequencies, $18$ MHz and $71$ MHz, we found that the
spectra have a small temperature dependence for the shift. This nature of
variation is expected since the absolute variation of the susceptibility
from $300$ K to $60$ K is itself very small. \ Figure $5$ shows variation of
NMR shift with temperature for LiCrO$_{2}$ above the region of the
transition temperature. The inset shows a plot of $K$ versus bulk
susceptibility $\chi $ with temperature as an implicit parameter. From this
plot, using Eq. (1), the hyperfine coupling constant A$_{hf}$ was estimated
to be $6$ $\pm 1$ kG /$\mu _{B}.$This result suggests a weak hyperfine
coupling between Li and Cr. This is in agreement with the results on NaCrO$%
_{2}$ (ref. 4).

Figure $6$ shows the spectra of LiCrO$_{2}$ in the region of transition.
Integrated spectral\ intensity, which is proportional to the number of $^{7}$%
Li nuclei, is roughly constant above $70$ K. At around $62$ K, intensity
falls sharply - as also shown in figure $7$ (corrected for temperature and $%
T_{2}$ effects). This sharp decline indicates the onset of long-range order.
The signal disappears completely at around $55$ K. In a conventional case,
the spectra broaden as one approaches the ordering temperature from above
and eventually disappear (from the window of observation) below the ordering
temperature due to the large static field that develops in the ordered
state. Here in LiCrO$_{2}$ at $1\ $T, the $^{7}$Li NMR signal reappeared at
around $35$ K. By around $12$ K, a small signal could be observed but the
spectrum was too broad to reliable obtain the lineshape using FT techniques.
Therefore, in figure $7$, data shown below $10$ K were taken by sweeping the
field at $18$ MHz. Resultant spectra are shown in the inset of figure $6$.
Here in LiCrO$_{2}$, unlike the case of NaCrO$_{2}$ (ref. 4), intensity is
not regained completely down to $4$ K. Spectra of Li$_{0.5}$CrO$_{2}$ also
showed a similar temperature dependence.

Spectra of LiCrO$_{2}$ in $4.3$ T showed a qualitatively similar behaviour
with small differences. The intensity sharply falls at around $63$ K and
continues to fall till about $52$ K. The signal is never completely lost and
about $30$ \% of the signal is regained by about $8$ K.

\subsubsection{ \textbf{Spin-lattice relaxation rate }$1/T_{1}$}

In LiCrO$_{2}$ and Li$_{0.5}$CrO$_{2}$ , for $H\approx 4.3$ T and $1$ T, the
recovery of the nuclear magnetisation after an inverting pulse was single
exponential. The recovery data fitted well to the expression,

$%
%TCIMACRO{%
%\FORMULA{(\frac{M(t)-M_{\infty }}{2M_{\infty }})=Ae^{(\frac{-t}{T_{1}})}+C}{(\frac{M(t)-M_{\infty }}{2M_{\infty }})=Ae^{(\frac{-t}{T_{1}})}+C}{evaluate}}%
%BeginExpansion
(\frac{M(t)-M_{\infty }}{2M_{\infty }})=Ae^{(\frac{-t}{T_{1}})}+C%
%EndExpansion
$

where \ $M(t)$ is the nuclear magnetisation at a time $t$ after an inverting
pulse and $M_{\infty }$ is the equilibrium value of the magnetisation.
Temperature dependence of 1/T$_{1}$ for $H\approx $ $1$ T for LiCrO$_{2}$ is
presented in figure $8$. In LiCrO$_{2\text{,}}$ the spin-lattice relaxation
rate shows a divergence around $63$ K due to slowing down of fluctuations as
one approaches magnetic order. This is around the same temperature at which
a broad maximum in the bulk susceptibility $\chi (T)$ was observed,
accompanied by an intensity loss of the NMR spectra. Below $59$ K, the $^{7}$%
Li signal was very weak and broad and did not allow for complete saturation
for accurate relaxation rate measurements.

At $71$ MHz ($H\approx $ $4.3$ T), relaxation rates were measured between $%
250$ K to $80$ K. \ The results were in reasonable agreement with those at $%
18$ MHz. Measurements on Li$_{0.5}$CrO$_{2}$ at $18$ MHz gave results
similar to those on LiCrO$_{2}$ .

\subsection{\textbf{Specific heat capacity}}

Results of the specific heat measurements are shown in figure $9$. Inset
shows the temperature dependence of specific heat capacity of LiCrO$_{2}$
and LiCoO$_{2}$. Layered LiCoO$_{2}$ is non-magnetic and isostructural with\
LiCrO$_{2}$. Also, in a Debye model picture, the Debye temperatures of the
two compounds are expected to be nearly the same taking into account their
molecular weights and unit cell volumes\cite{thetaD explanation}. Indeed, at
high temperatures, the heat capacities of the two compounds coincide. The
magnetic contribution to the specific heat of LiCrO$_{2}$ (C$_{mag}$) was
determined by subtracting the specific heat of LiCoO$_{2}$ from that of LiCrO%
$_{2}$. C$_{mag}$ shows a peak at around $63$ K, in good agreement with
anomalies observed by us in bulk magnetic susceptibility and NMR
measurements.

\section{\textbf{DISCUSSION}}

The magnetic susceptibility of LiCrO$_{2}$ has been reported by several
groups.\cite{licro2 tauber,licro2 delmas,licro2 moreno} \ The main
conclusion has been that a Curie-Weiss behaviour (indicative of
non-interacting local moments) is valid only above about $400$ K. \ A large
range from $400$ K to $63$ K (at which long-range order finally sets in) in
which magnetic correlations are present is probably due to the frustrating
magnetic interactions of the triangular Cr planes.

We found that delithiation of LiCrO$_{2}$ did not have a significant impact
on its magnetic properties. \ This is quite unlike the case of NaCoO$_{2}$
where the magnetic ground state depends strongly on the Na content. \ This
implies that the smaller Li ion does not impact the magnetism of the Cr
layers nor does it contribute mobile charge carriers in the planes. \ 

As seen from the structure, each Li has a super-transferred hyperfine
coupling to the Cr layers \textit{via} oxygen ions. Since seven Cr ions
(three ions from three triangles of which two ions are common) from each
layer might be expected to affect each Li, the Li senses a distribution of
susceptibilities and hence a linewidth which increases with decreasing
temperature as also with increasing field. In the ordered state, a $120%
%TCIMACRO{\U{b0}}%
%BeginExpansion
{{}^\circ}%
%EndExpansion
$ alignment of the spins on the vertices of the triangles will give a
cancellation of hyperfine fields at the Li site.The linewidth might remain
large due to any structural and magnetic disorder. In the case of NaCrO$_{2}$%
, an extended fluctuating regime was reported below $T_{\text{N}}$ and the
full NMR\ signal intensity was recovered by about $10$ K. \ While Olariu 
\textit{et al}.\cite{nacro2 mendels} did not show NMR $1$/$T_{1}$ data for
NaCrO$_{2}$, their relaxation rate data from $\mu $SR showed a peak at $30$
K which is much lower than the ordering temperature of $40$ K in that
compound. \ They therefore concluded that an extended fluctuating regime
exists for NaCrO$_{2}$ below its $T_{\text{N}}$. \ On the other hand, our $%
^{7}$Li NMR spin-lattice relaxation rate for LiCrO$_{2}$ diverges at $T_{%
\text{N}}$. \ Also, preliminary $\mu $SR\ measurements\cite{muSR licro2} on
our samples of LiCrO$_{2}$ have indicated the absence of an extended
fluctuation regime below $T_{\text{N}}$.\ A stronger interplane interaction
in LiCrO$_{2}$ as compared to NaCrO$_{2}$ is also indicated.

Theoretical study by Kawamura \textit{et al}.\cite{kawamura kt} have
suggested the occurrence of Kosterlitz-Thouless-Berezinskii type of phase
transition\cite{kosterlitz1,kosterlitz3,kosterlitz2} in $2D$ triangular
lattice antiferromagnets, complimented by the inherent frustration effects.
Here, the correlation between the spins generate vortices which stay as
bounded pairs at low temperatures. But at a critical temperature,$T_{\text{%
KTB}}$ ($\neq 0$), these bounded vortex-antivortex pairs dissociate to free
vortices resulting in a topological phase transition. Spin-lattice
relaxation measurements can provide clues for the order driven by the free
vortices at temperatures above $T_{\text{KTB}}$ in $2D$ antiferromagnets\cite%
{gaveau kosterlitz,mertens kosterlitz}. For $T>T_{\text{KTB}}$, spin-lattice
relaxation rate $1/T_{\text{1}}$ has two contributions: The first due to
free vortices and the second due to spin-wave excitations, as given in the
equation below.

\begin{equation}
\frac{1}{T_{1}}=\frac{(\gamma A_{hf}/2\pi )^{2}}{\sqrt{\pi .n_{v}}U}+\alpha
T+\beta
\end{equation}

where $A_{hf}$ is the hyperfine coupling constant, $\gamma $ is the
gyromagnetic ratio for $^{7}$Li, and corresponding to a $2D$ correlation
length of $\xi ,$ $n_{v}=1/(2\xi )^{2}$ is the density of free vortices\cite%
{kosterlitz nv}:

\begin{equation}
n_{v}=\left( \frac{1}{2\xi _{0}}\right) ^{2}\exp \left( \frac{-2b}{\sqrt{%
\tau }}\right)
\end{equation}

where, $\tau =\left( T/T_{\text{KTB}}\right) -1$ and $\xi _{0}\simeq a,$ the
lattice parameter and $b$ is a scaling parameter. The ideal value for $b$ is 
$\pi /2$ , but theoretical calculation and later experimental reports\cite%
{mertens kosterlitz,gaveau kosterlitz,kosterlitz heinrich} allow for smaller
values. Free vortices move through the lattice inducing spin flipping and
the vortex velocity $U$ (ref. 24) is given by

\begin{equation}
U=\left[ \frac{\pi }{2}\left( \frac{JS^{2}a^{2}}{\hbar }\right) ^{2}n_{v}\ln 
\frac{k_{B}T_{KTB}}{JS^{2}n_{v}a^{2}}\right] ^{1/2}
\end{equation}

The solid-line in figure $8$ shows fit of Eq. (2) to our 1/T$_{1}$(T) data,
assuming $A_{hf}=6$ kG $/$ $\mu _{B}$ (from the $K$ vs $\chi $ plot- Fig.
5), $b=0.86$ and $J/k_{B}=40$ K (from the report by Delmas \textit{et al.}%
\cite{licro2 delmas}). The corresponding $T_{\text{KTB}}$ for LiCrO$_{2}$
was found to be $55$ K which is close to the value obtained by Ajiro \textit{%
et al.}\cite{ajiro kosterlitz} of $60$ K and might explain the temperature
dependence of the spin-lattice relaxation rate above $T_{\text{N}}$.

The coefficient of the linear term in Eq. (2) was found to be, $\alpha =0.45$
sec$^{-1}$K$^{-1}.$ As mentioned earlier, the linear increase in 1/T$_{1}$
data at temperatures above $120$ K can be attributed to spin-wave
excitations. But in the paramagnetic ( high temperature) limit, one expects
the relaxation rate value to level off - as the fluctuations of local
moments is fast and random. In a case where the nuclear relaxation mechanism
is mainly governed by fluctuations of localised spins, $1/T_{1}$(T) in the
high temperature limit, 1/T$_{1\infty }$ is given by\cite%
{1/T1infinity1,1/T1infinity2}

\begin{equation}
\frac{1}{T_{1\infty }}=\sqrt{2\pi }\left( \frac{\gamma g\mu _{B}A_{hf}}{%
z^{\prime }}\right) ^{2}\frac{z^{\prime }S(S+1)}{3\omega _{ex}}
\end{equation}

where $\omega _{ex}$ is the exchange frequency of local moments given by $%
\omega _{ex}=k_{B}\left\vert \theta _{CW}\right\vert /\left[ \text{%
%TCIMACRO{\U{127}}%
%BeginExpansion
h{\hskip-.2em}\llap{\protect\rule[1.1ex]{.325em}{.1ex}}{\hskip.2em}%
%EndExpansion
}\sqrt{zS(S+1)/6}\right] .$ $z$ and $z^{\prime }$ are the number of local
moments exchange coupled with each other ($=6,$ here) and that of local
moments interacting with the probing nuclei ($=6,$ here) at non-magnetic
crystal sites, respectively. The other parameters represent usual physical
constants. For LiCrO$_{2}$ , $g=1.98$, was taken from reports of EPR
measurements by Moreno \textit{et} \textit{al.}\cite{licro2 moreno}. Thus,
using Eq. (5), spin-lattice relaxation rate of LiCrO$_{2}$ in the
paramagnetic region was found to be $190$ sec$^{-1}$. The linear increase in
1/T$_{1}$, with temperature, observed by us above $120$ K will yield a value
of $190$ sec$^{-1}$ at about $470$ K; $i$.$e$. LiCrO$_{2}$ seems to reach
the paramagnetic limit only at $470$ K. This finding is in agreement with $%
\chi (T)$ data\cite{licro2 tauber,licro2 delmas,licro2 moreno} which found $%
\left\vert \theta _{CW}\right\vert \sim 570-700$ K$.$

The heat capacity of non-magnetic LiCoO$_{2}$ could be fit reasonably well
in the full temperature range using the Debye model which yields for the
specific heat at constant volume $C_{\text{v}}$

\begin{equation}
C_{\text{v}}=9rN_{A}k_{B}[(\frac{4T^{3}}{\theta ^{3}})\int_{0}^{\theta /T}%
\frac{x^{3}dx}{e^{x}-1}-\frac{\theta /T}{e^{\theta /T}-1}]
\end{equation}%
where $r$ is the number of atoms per formula unit, $N_{A}$ is the Avogadro's
number and $k_{B}$ is the Boltzmann constant, and $\theta $ is the Debye
temperature. For LiCoO$_{2}$, $\theta $ was found to be about $800$ K.

The total entropy related to magnetic order in LiCrO$_{2}$ can be determined
by finding the area under the curve of the $C_{mag}$/$T$ vs $T$ plot of
figure $9$. \ We obtained a value of $11.65$ J/mol-K for the magnetic
entropy $S_{mag}$ which is in reasonable agreement with the expected value $%
R\ln (2S+1)=11.53$ J/mol-K where $R$ is the gas constant and $S=3/2$ for Cr$%
^{3+}$. \ Nearly $80$ \% of the contribution to the magnetic entropy change
for LiCrO$_{2}$ comes from temperatures above $T_{\text{N}}$. This appears
similar to the case of NaCrO$_{2}$ but in contrast to the entropy behaviour
in conventional long-range ordered materials.

Below $240$ K, Li$_{x}$CrO$_{2}$ shows a small history dependence
(difference between the ZFC and FC susceptibility curves). This ZFC-FC
splitting might be connected to the formation of frozen magnetic clusters,
which are singular regions with quite local magnetic interactions. Such a
history dependence at temperatures much above the transition temperature has
been reported earlier in geometrically frustrated compounds like Li$_{x}$Zn$%
_{1-x}$V$_{2}$O$_{4}$ $(0\leq x\leq 0.9)$ $\left[ \text{ref. 28,29}\right] $
and Tb$_{2}$Mo$_{2}$O$_{7}$ $\left[ \text{ref. 30}\right] $. Here in LiCrO$%
_{2}$, the origin of the clusters may be from the complex intra-layer
magnetic correlations additionally influenced by the possible presence of
free vortices.

The measurements show that LiCrO$_{2}$ is long range ordered below $62$ K.
The bulk magnetic susceptibility and magnetic specific heat, $C_{mag}$
suggests the presence of a short-range order in LiCrO$_{2}$ even at
temperatures well above $62$ K. Susceptibility data show a relatively weak
anomaly at the ordering temperature. NMR spin-lattice relaxation data
suggests the presence of free vortices in the lattice above $62$ K. From our 
$^{7}$Li NMR relaxation rate analysis, based on expectations for a pure
Kosterlitz-Thouless-Berezinskii system, we determined $T_{\text{KTB}}$ = $55$
K. In the present system, however, inter-plane interactions lead to
long-range order below $62$ K.

\ 

\section{CONCLUSION}

In summary, we have performed $^{7}$Li NMR, bulk susceptibility, and heat
capacity measurements on the TLHAF LiCrO$_{2}$. Anomalies in the NMR
intensity, spin-lattice relaxation rate, bulk susceptibility, and heat
capacity indicate an antiferromagnetic ordering temperature ($T_{\text{N}}$)
of about $62$ K in LiCrO$_{2}$. An extended fluctuating regime below $T_{%
\text{N}}$ appears to be much less prominent in LiCrO$_{2}$ compared to NaCrO%
$_{2}$. The interplanar magnetic interactions, as well, are therefore
stronger in LiCrO$_{2}$ compared to NaCrO$_{2}$. A significant magnetic
contribution to the specific heat exists above $T_{\text{N}}$. Magnetic
susceptibility and specific heat data suggest presence of considerable
amount of magnetic short-range order even at temperatures well above $T_{%
\text{N}}$.This is corroborated by the temperature dependence of the $^{7}$%
Li nuclear spin-lattice relaxation rate 1/T$_{1}$(T) above $T_{\text{N}}$
which was modeled to arise from free moving vortices in the LiCrO$_{2}$
lattice.\ Our value of $T_{\text{KTB}}$ = $55$ K is similar to that
estimated by Ajiro \textit{et al.}\cite{ajiro kosterlitz} of $60$ K. Li
depletion does not have a significant effect on the magnetic properties of
LiCrO$_{2}$ in contrast to isostructural NaCoO$_{2}$. More experiments and
theoretical investigations on TLHAF's are needed to obtain a full
understanding of their static and dynamic magnetic properties.

\begin{acknowledgments}
This work was supported by the Deutsche Forschungsgemeinschaft via the
Sonderforschungsbereich 484 (Augsburg) and partly by BMBF via VDI/EKM, FKZ
13N6917. We acknowledge helpful discussions with P. Mendels.
\end{acknowledgments}

\bigskip

\textbf{Figure Captions}

FIG. 1 (Color online). XRD patterns of (a) LiCrO$_{2}$ and (b) Li$_{0.5}$CrO$%
_{2}$. Both the peak sets could be completely indexed using the space group $%
\mathit{R}\overline{3}m.$

FIG. 2 (Color online). A schematic of the crystal structure of LiCrO$_{2}$.

FIG. 3 (Color online). Temperature dependence of bulk magnetic
susceptibility $(M/H)$ of LiCrO$_{2}$ and Li$_{0.5}$CrO$_{2}$ in an applied
field of $1$ T between $2$ K to $300$ K.

FIG. 4 (Color online). Zero-field-cooled (ZFC)and field-cooled (FC) magnetic
susceptibilities for LiCrO$_{2}$ as a function of temperature between $2$ K
and $300$ K in a field of $50$ G. The plot shows a clear splitting of the
ZFC and FC curves. The scatter in the data is due to the small magnetisation
produced by a $50$ G field which is near the limiting sensitivity of the VSM.

FIG. 5 (Color online). Variation of $^{7}$Li NMR shift, $K(T)$ with
temperature for LiCrO$_{2}$ above the region of transition temperature.
Inset shows plot of $K$ versus bulk susceptibility, $\chi $ with temperature
as an implicit parameter. From this plot, using Eq. (1), hyperfine coupling
constant was estimated to be nearly $6$ kG /$\mu _{B}.$.

FIG. 6 (Color online). $^{7}$Li NMR FFT spectra of LiCrO$_{2}$ in the region
of transition. Inset shows field sweep spectra at low temperatures $10$ K
and $4$ K.

FIG. 7 (Color online). Temperature dependence of the $^{7}$Li NMR\ spectral
intensity of LiCrO$_{2}$ obtained at $18$ MHz between $4$ K and $90$ K.
Here, echo integrated intensities have been corrected for temperature and $%
T_{2}$ effects. The error bar shown at $73$ K is exemplary and indicates the
value of error common to all the data points shown above $58$ K. NMR\ signal
was completely absent from $56$ K to $35$ K. The low-temperature points at $%
10$ K and $4$ K were obtained from the field- sweep spectra ($cf$. the inset
of Fig. $6$).

FIG. 8 (Color online). Temperature dependence of $^{7}$Li NMR spin-lattice
relaxation rate, $1$/$T_{1}$ for LiCrO$_{2}$ at $H=1.08$ T. Solid line shown
in the plot is fit of the relaxation data with Eq. (2) (discussed in section
IV).

FIG. 9 (Color online). Variation of magnetic specific heat divided by
temperature (C$_{mag}$/$T$) for LiCrO$_{2}$. C$_{mag}$ was obtained by
subtracting the specific heat of non-magnetic LiCoO$_{2}$ (see text). Inset:
Specific heat capacity versus temperature for LiCrO$_{2}$ and LiCoO$_{2}$.

\ \ \ \ \ \ \ \ \ \ \ \ \ \ \ \ \ \ \ \ \ \ \ \ \ \ \ \ \ \ \ \ \ \ \ \ \ \
\ \ \ \ \ \ \ \ \ \ \ \ \ \ \ \ \ \ \ \ \ \ \ \ \ \ \ \ \ \ \ \ \ \ \ \ \ \
\ \ \ \ \ \ \ \ \ \ \ \ \ \ \ \ \ \ \ \ \ \ \ \ \ \ \ \ \ \ \ \ \ \ \ \ \ \
\ \ \ \ \ \ \ \ \ \ \ \ \ \ \ \ \ \ \ \ \ \ \ \ \ \ \ \ \ \ \ \ \ \ \ \ \ \
\ \ \ \ \ \ \ \ \ \ \ \ \ \ \ \ \ \ \ \ \ \ \ \ \ \ \ \ \ \ \ \ \ \ \ \ \ \
\ \ \ \ \ \ \ \ \ \ \ \ \ \ \ \ \ \ \ \ \ \ \ \ \ \ \ \ \ \ \ \ \ \ \ \ \ \
\ \ \ \ \ \ \ \ \ \ \ \ \ \ \ \ \ \ \ \ \ \ \ \ \ \ \ \ \ \ \ \ \ \ \ \ \ \
\ \ \ \ \ \ \ \ \ \ \ \ \ \ \ \ \ \ \ \ \ \ \ \ \ \ \ \ \ \ \ \ \ \ \ \ \ \
\ \ \ \ \ \ \ \ \ \ \ \ \ \ \ \ \ \ \ \ \ \ \ \ \ \ \ \ \ \ \ \ \ \ \ \ \ \
\ \ \ \ \ \ \ \ \ \ \ \ \ \ \ \ \ \ \ \ \ \ \ \ \ \ \ \ \ \ \ \ \ \ \ \ \ \
\ \ \ \ \ \ \ \ \ \ \ \ \ \ \ \ \ \ \ \ \ \ \ \ \ \ \ \ \ \ \ \ \ \ \ \ \ \
\ \ \ \ \ \ \ \ \ \ \ \ \ \ \ \ \ \ \ \ \ \ \ \ \ \ \ \ \ \ \ \ \ \ \ \ \ \
\ \ \ \ \ \ \ \ \ \ \ \ \ \ \ \ \ \ \ \ \ \ \ \ \ \ \ \ \ \ \ \ \ \ \ \ \ \
\ \ \ \ \ \ \ \ \ \ \ \ \ \ \ \ \ \ \ \ \ \ \ \ \ \ \ \ \ \ \ \ \ \ \ \ \ \
\ \ \ \ \ \ \ \ \ \ \ \ \ \ \ \ \ \ \ \ \ \ \ \ \ \ \ \ \ \ \ \ \ \ \ \ \ \
\ \ \ \ \ \ \ \ \ \ \ \ \ \ \ \ \ \ \ \ \ \ \ \ \ \ \ \ \ \ \ \ \ \ \ \ \ \
\ \ \ \ \ \ \ \ \ \ \ \ \ \ \ \ \ \ \ \ \ \ \ \ \ \ \ \ \ \ \ \ \ \ \ \ \ \
\ \ \ \ \ \ \ \ \ \ \ \ \ \ \ \ \ \ \ \ \ \ \ \ \ \ \ \ \ \ \ \ \ \ \ \ \ \
\ \ \ \ \ \ \ \ \ \ \ \ \ \ \ \ \ \ \ \ \ \ \ \ \ \ \ \ \ \ \ \ \ \ \ \ \ \
\ \ \ \ \ \ \ \ \ \ \ \ \ \ \ \ \ \ \ \ \ \ \ \ \ \ \ \ \ \ \ \ \ \ \ \ \ \
\ \ \ \ \ \ \ \ \ \ \ \ \ \ \ \ \ \ \ \ \ \ \ \ \ \ \ \ \ \ \ \ \ \ \ \ \ \
\ \ \ \ \ \ \ \ \ \ \ \ \ \ \ \ \ \ \ \ \ \ \ \ \ \ \ \ \ \ \ \ \ \ \ \ \ \
\ \ \ \ \ \ \ \ \ \ \ \ \ \ \ \ \ \ \ \ \ \ \ \ \ \ \ \ \ \ \ \ \ \ \ \ \ \
\ \ \ \ \ \ \ \ \ \ \ \ \ \ \ \ \ \ \ \ \ \ \ \ \ \ \ \ \ \ \ \ \ \ \ \ \ \
\ \ \ \ \ \ \ \ \ \ \ \ \ \ \ \ \ \ \ \ \ \ \ \ \ \ \ \ \ \ \ \ \ \ \ \ \ \
\ \ \ \ \ \ \ \ \ \ \ \ \ \ \ \ \ \ \ \ \ \ \ \ \ \ \ \ \ \ \ \ \ \ \ \ \ \
\ \ \ \ \ \ \ \ \ \ \ \ \ \ \ \ \ \ \ \ \ \ \ \ \ \ \ \ \ \ \ \ \ \ \ \ \ \
\ \ \ \ \ \ \ \ \ \ \ \ \ \ \ \ \ \ \ \ \ \ \ \ \ \ \ \ \ \ \ \ \ \ \ \ \ \
\ \ \ \ \ \ \ \ \ \ \ \ \ \ \ \ \ \ \ \ \ \ \ \ \ \ \ \ \ \ \ \ \ \ \ \ \ \
\ \ \ \ \ \ \ \ \ \ \ \ \ \ \ \ \ \ \ \ \ \ \ \ \ \ \ \ \ \ \ \ \ \ \ \ \ \
\ \ \ \ \ \ \ \ \ \ \ \ \ \ \ \ \ \ \ \ \ \ \ \ \ \ \ \ \ \ \ \ \ \ \ \ \ \
\ \ \ \ \ \ \ \ \ \ \ \ \ \ \ \ \ \ \ \ \ \ \ \ \ \ \ \ \ \ \ \ \ \ \ \ \ \
\ \ \ \ \ \ \ \ \ \ \ \ \ \ \ \ \ \ \ \ \ \ \ \ \ \ \ \ \ \ \ \ \ \ \ \ \ \
\ \ \ \ \ \ \ \ \ \ \ \ \ \ \ \ \ \ \ \ \ \ \ \ \ \ \ \ \ \ \ \ \ \ \ \ \ \
\ \ \ \ \ \ \ \ \ \ \ \ \ \ \ \ \ \ \ \ \ \ \ \ \ \ \ \ \ \ \ \ \ \ \ \ \ \
\ \ \ \ \ \ \ \ \ \ \ \ \ \ \ \ \ \ \ \ \ \ \ \ \ \ \ \ \ \ \ \ \ \ \ \ \ \
\ \ \ \ \ \ \ \ \ \ \ \ \ \ \ \ \ \ \ \ \ \ \ \ \ \ \ \ \ \ \ \ \ \ \ \ \ \
\ \ \ \ \ \ \ \ \ \ \ \ \ \ \ \ \ \ \ \ \ \ \ \ \ \ \ \ \ \ \ \ \ \ \ \ \ \
\ \ \ \ \ \ \ \ \ \ \ \ \ \ \ \ \ \ \ \ \ \ \ \ \ \ \ \ \ \ \ \ \ \ \ \ \ \
\ \ \ \ \ \ \ \ \ \ \ \ \ \ \ \ \ \ \ \ \ \ \ \ \ \ \ \ \ \ \ \ \ \ \ \ \ \
\ \ \ \ \ \ \ \ \ \ \ \ \ \ \ \ \ \ \ \ \ \ \ \ \ \ \ \ \ \ \ \ \ \ \ \ \ \
\ \ \ \ \ \ \ \ \ \ \ \ \ \ \ \ \ \ \ \ \ \ \ \ \ \ \ \ \ \ \ \ \ \ \ \ \ \
\ \ \ \ \ \ \ \ \ \ \ \ \ \ \ \ \ \ \ \ \ \ \ \ \ \ \ \ \ \ \ \ \ \ \ \ \ \
\ \ \ \ \ \ \ \ \ \ \ \ \ \ \ \ \ \ \ \ \ \ \ \ \ \ \ \ \ \ \ \ \ \ \ \ \ \
\ \ \ \ \ \ \ \ \ \ \ \ \ \ \ \ \ \ \ \ \ \ \ \ \ \ \ \ \ \ \ \ \ \ \ \ \ \
\ \ \ \ \ \ \ \ \ \ \ \ \ \ \ \ \ \ \ \ \ \ \ \ \ \ \ \ \ \ \ \ \ \ \ \ \ \
\ \ \ \ \ \ \ \ \ \ \ \ \ \ \ \ \ \ \ \ \ \ \ \ \ \ \ \ \ \ \ \ \ \ \ \ \ \
\ \ \ \ \ \ \ \ \ \ \ \ \ \ \ \ \ \ \ \ \ \ \ \ \ \ \ \ \ \ \ \ \ \ \ \ \ \
\ \ \ \ \ \ \ \ \ \ \ \ \ \ \ \ \ \ \ \ \ \ \ \ \ \ \ \ \ \ \ \ \ \ \ \ \ \
\ \ \ \ \ \ \ \ \ \ \ \ \ \ \ \ \ \ \ \ \ \ \ \ \ \ \ \ \ \ \ \ \ \ \ \ \ \
\ \ \ \ \ \ \ \ \ \ \ \ \ \ \ \ \ \ \ \ \ \ \ \ \ \ \ \ \ \ \ \ \ \ \ \ \ \
\ \ \ \ \ \ \ \ \ \ \ \ \ \ \ \ \ \ \ \ \ \ \ \ \ \ \ \ \ \ \ \ \ \ \ \ \ \
\ \ \ \ \ \ \ \ \ \ \ \ \ \ \ \ \ \ \ \ \ \ \ \ \ \ \ \ \ \ \ \ \ \ \ \ \ \
\ \ \ \ \ \ \ \ \ \ \ \ \ \ \ \ \ \ \ \ \ \ \ \ \ \ \ \ \ \ \ \ \ \ \ \ \ \
\ \ \ \ \ \ \ \ \ \ \ \ \ \ \ \ \ \ \ \ \ \ \ \ \ \ \ \ \ \ \ \ \ \ \ \ \ \
\ \ \ \ \ \ \ \ \ \ \ \ \ \ \ \ \ \ \ \ \ \ \ \ \ \ \ \ \ \ \ \ \ \ \ \ \ \
\ \ \ \ \ \ \ \ \ \ \ \ \ \ \ \ \ \ \ \ \ \ \ \ \ \ \ \ \ \ \ \ \ \ \ \ \ \
\ \ \ \ \ \ \ \ \ \ \ \ \ \ \ \ \ \ \ \ \ \ \ \ \ \ \ \ \ \ \ \ \ \ \ \ \ \
\ \ \ \ \ \ \ \ \ \ \ \ \ \ \ \ \ \ \ \ \ \ \ \ \ \ \ \ \ \ 


\begin{thebibliography}{99}
\bibitem{thetaD explanation} Debye temperature, $\theta $ $\propto 1/\left(
V^{1/3}M^{1/2}\right) $ where $V$ is the volume of the crystal unit cell and 
$M$ is the molecular mass. Here in our case, higher molecular mass of LiCoO$%
_{2}$ is almost neutralised by the bigger volume of LiCrO$_{2}$ to get an
expected ratio of Debye temperatures to be about 1.01.
\end{thebibliography}
\end{document}